\documentclass{aa}
\usepackage[varg]{txfonts}
\usepackage{wrapfig}
\usepackage[export]{adjustbox}
\begin{document}

\title{Refining Saturn's deuterium-hydrogen ratio via IRTF/TEXES spectroscopy}

\author{James S.D. Blake (1), Leigh N. Fletcher (1), Thomas K. Greathouse (2), Glenn S. Orton (3), Henrik Melin (1), Mike T. Roman (1), Arrate Antuñano (1), Padraig T. Donnelly (1), Naomi Rowe-Gurney (1), Oliver King (1) 
}
\offprints{J.S.D. Blake, \email{jsdb3@le.ac.uk}}

\institute{Department of Physics \& Astronomy, University of Leicester, University Road, Leicester, LE1 7RH, UK, 
\and Southwest Research Institute, San Antonio, Texas, TX, USA
\and Jet Propulsion Laboratory, California Institute of Technology, 4800 Oak Grove Drive, Pasadena, CA, 91109, USA}

\date{Received 3 July 2020 / Accepted 25 January 2021}

\abstract{
The abundance of deuterium in giant planet atmospheres provides constraints on the reservoirs of ices incorporated into these worlds during their formation and evolution. Motivated by discrepancies in the measured deuterium-hydrogen ratio (D/H) on Jupiter and Saturn, we present a new measurement of the D/H ratio in methane for Saturn from ground-based measurements. We analysed a spectral cube (covering 1151-1160 cm$^{-1}$ from 6 February 2013) from the Texas Echelon Cross Echelle Spectrograph (TEXES) on NASA’s Infrared Telescope Facility (IRTF) where emission lines from both methane and deuterated methane are well resolved. Our estimate of the D/H ratio in stratospheric methane, $1.65_{-0.21}^{+0.27} \times 10^{-5}$ is in agreement with results derived from Cassini CIRS and ISO/SWS observations, confirming the unexpectedly low CH$_{3}$D abundance. Assuming a fractionation factor of $1.34 \pm 0.19$ we derive a hydrogen D/H of $1.23_{-0.23}^{+0.27} \times 10^{-5}$. This value remains lower than previous tropospheric hydrogen D/H measurements of (i) Saturn $2.10  (\pm 0.13) \times 10^{-5}$, (ii) Jupiter $2.6 (\pm 0.7) \times 10^{-5}$ and (iii) the proto-solar hydrogen D/H of  $2.1 (\pm 0.5) \times 10^{-5}$, suggesting that the fractionation factor may not be appropriate for stratospheric methane, or that the D/H ratio in Saturn’s stratosphere is not representative of the bulk of the planet.}


\titlerunning{Refining Saturn's D-H Ratio}
\authorrunning{J.S.D Blake et al.}
\maketitle

\section{Introduction}
\label{intro}

\begin{table*}[t]

\centering                                     
\begin{tabular}{c c c c p{3cm} p{4cm}}          
\hline\hline                        
\multicolumn{2}{c}{Saturn} & \multicolumn{2}{c}{Jupiter} & References & Instrument/Model \\    
\hline                                   
Hydrogen &	Methane	& Hydrogen	& Methane & &	\\
\hline
 - & $1.7_{-0.8}^{+1.7}$ & - & - & Owen et al (1986) & Model – Fourier Transform instrument, Kitt Peak National Observatory\\       
 - & $1.7(\pm1.1)$ & - & - & Noll and Larson (1991) & Model – Fourier Transform Spectrometer, IRTF\\       
 $2.3_{-0.8}^{+1.2}$ & - & - & - & Griffin et al. (1996) & ISO – LWS\\
 $1.85_{-0.6}^{+0.85}$ (m) & $2.0_{-0.7}^{+1.4}$ (m) & $2.4 \pm 0.4$ (m) & 	$2.2 \pm 0.7$ (m) & Lellouch et al (2001) & ISO - SWS\\     
- & $2.4 \pm 0.5$ & - & - & Bézard et al. (2003) & TEXES\\   
- & - & - & $2.4 \pm 0.7$ & Bjoraker (1986) & Voyager IRIS\\   
- & $1.6 \pm 0.2$ & - & - & Fletcher et al. (2009b) & CIRS\\
$2.1 \pm 0.13$ (m) & $2.82_{-0.55}^{+0.6}$ (i) & $2.95 \pm 0.55$ & - & Pierel et al. (2017) & CIRS\\
$1.62 \pm 0.61$ & - & - & - &  & ISO - SWS\\
- & - & $2.6 \pm 0.7$ & - & Mahaffy et al. (1998) & Galileo\\
$1.23_{-0.23}^{+0.27}$ (i) & $1.65_{-0.21}^{+0.27}$ (m) & - & - & This work & TEXES\\     
\hline                                             
\end{tabular}
\caption{The quoted D/H ratios are of the order $\times 10^{-5}$. Entries marked (i) are inferred from corresponding measurements marked (m) for Lellouch et al. (2001), Pierel et al. (2017) and this work. The values quoted for the Methane column assume a fractionation factor of $1.34 \pm 0.19$ (Lellouch et al. 2001).}              
\label{table:1}      
\end{table*}

The abundance of deuterium in planetary bodies has long been a focus of research owing to its usefulness in constraining models of planet formation and evolution (Hersant et al. 2001). The deuterium-hydrogen ratio (D/H) is known to increase with heliocentric distance due to the fractionation of deuterium in the formation of icy grains. As such, measuring the D/H ratio in planetary atmospheres and ice-rich bodies (moons and comets) as a function of heliocentric distance helps to constrain the distribution of ices in the early Solar System. Molecular hydrogen constitutes the most abundant source of hydrogen in the atmosphere (around 85\% by volume) and has been used previously to determine the HD/H$_{2}$ ratio in planetary tropospheres using far-infrared spectra (Feuchtgruber et al., 2013). The D/H ratio in planetary stratospheres can also be assessed using a mid-IR spectral setting that contains both methane (CH$_{4}$) and deuterated methane (CH$_{3}$D) emission lines as described in Section 2.

There are currently no known methods for the natural production of deuterium, with the exception of the formation of the Universe. There is however a natural loss mechanism, the nucleosynthesis of helium in stars, including the Sun (Griffin et al., 1996, Ferlet \& Lemoine, 1998, Lellouch et al., 2001). Thus, the Solar System D/H ratio in H$_2$ has been decreasing from the protosolar value of $2.1 (\pm 0.5) \times 10^{-5}$ (Geiss and Gloeckler, 1998) to the local interstellar medium value of $1.5 (\pm 0.1) \times 10^{-5}$ throughout time. We therefore expect the D/H ratio of Jupiter and Saturn to be representative of the protosolar value incorporated into the icy building blocks during planetary formation. Several previous measurements of the D/H have been made for each of the giant planets, a sample of which is shown in Table 1. The D/H ratio for the two gas giants is smaller than that measured for the ice giants Uranus and Neptune - Feuchtgruber et al (2013) report a hydrogen D/H of $4.4 (\pm 0.4) \times 10^{-5}$ for Uranus and $4.1 (\pm 0.4) \times 10^{-5}$ for Neptune. Independent near-infrared and mid-infrared analyses of Neptune by Irwin et al. (2014) and Fletcher et al (2010) report a CH$_{3}$D/CH$_{4}$ ratio of $3.0 (\pm 1.0) \times 10^{-4}$, which must be divided by 4 to give a (D/H)$_{CH_{4}}$ = $7.5 (\pm2.5)\times 10^{-5}$ in Neptune’s atmosphere.  Fletcher et al. (2010) assumed an old isotopic fractionation factor f=(D/H)$_{CH_{4}}$/(D/H)$_{H_{2}}$=1.25 (Fegley and Prinn, 1988) to give an estimate of Neptune’s hydrogen (D/H)$_{H_{2}}$ of $6 (\pm2) \times 10^{-5}$ in agreement with ISO/SWS analysis by Feuchtgruber et al. (1999).  However, if we use a more suitable Neptunian $f=1.6\pm0.2$ from Lecluse et al. (1996), this would give a (D/H)$_{H_{2}}$ of $4.7 (\pm1.7) \times 10^{-5}$ that is more consistent with the Herschel/PACS analysis of Feuchtgruber et al. (2013). Comparing the measured (D/H)$_{CH_{4}}$ and (D/H)$_{H_{2}}$ for Neptune, this implies a fractionation factor of $1.8\pm0.6$, closer to the estimate of $f=1.6\pm0.2$ (Lecluse et al., 1996).  These values are consistent with the Ice Giants accreting greater quantities of ices (compared to hydrogen gas) during their formation.

The challenge of measuring the D/H ratio in a planetary atmosphere is considerable, even with a visiting spacecraft like Cassini. It is notable that CIRS estimates of Saturn’s D/H in methane, inferred from HD lines in the far-infrared by Pierel et al. (2017) and measured in CH$_{3}$D by Fletcher et al. (2009b), are both in disagreement with each other. Taken at face value, the D/H in stratospheric methane from Fletcher et al. (2009b) (mid-infrared) would appear to be smaller than the D/H in hydrogen (far infrared), which is both unexpected and hard to explain. Furthermore, both estimates are smaller than the proto-solar value of $2.1 (\pm0.5) \times 10^{-5}$, as well as being smaller than the estimates for Jupiter (e.g. the directly measured Galileo probe result of Mahaffy et al., 1998). It is expected that the D/H in methane should be higher than in hydrogen, further, it should be higher than values at Jupiter due to deuterium enrichment with heliocentric distance. No explanation is currently offered for this phenomenon as its existence has always been disputed (Bézard et al., 2003). These discrepancies call into question the D/H measurements in Saturn’s atmosphere, and prompt us to revisit them in this study. Table 1 shows that measurements from higher spectral resolution spectroscopic instruments such as ISO, which yielded a D/H ratio in methane of $2.0 \times 10^{-5}$ from Lellouch et al. (2001), are more consistent with one another than lower spectral resolution measurements such as those from Cassini CIRS. The D/H of Saturn has been noted to be lower than the D/H of Jupiter (Pierel et al., 2017), even when considering error margins. Motivated by these discrepancies with the Cassini results, we seek to re-examine Saturn’s D/H ratio using high-resolution ground-based spectroscopy.

We build upon the previous work using the Texas Echelon cross Echelle Spectrograph (TEXES) instrument (Lacy et al., 2002) on the NASA Infrared Telescope Facility (IRTF). TEXES is currently the world-leading instrument for high-spectral resolution mid-infrared observations (maximum R = 100,000) and therefore has the capacity to produce the most accurate measurement of D/H from a ground-based telescope. Prior work on Saturn using this instrument has included the first ever measurement of propane at Saturn (Greathouse et al., 2005), analysis of nitrogen isotopologues in both Saturn’s and Jupiter’s ammonia reservoirs (Fletcher et al., 2014), as well as exploring the stratospheric aftermath of the 2010 storm observed on Saturn (Fouchet et al., 2016). This work uses the same archive of TEXES observations collected at the IRTF to examine the D/H ratio of methane in the 8.6 $\mu$m region. 

\section{Observation and data reduction}
\label{obs}

We observed Saturn in the N-band 8.6 $\mu$m region for one night in February 2013, using the high spectral resolution of the TEXES instrument and the 3-meter diameter mirror of NASA’s IRTF. On 6 February 2013, Saturn’s angular diameter was 17.2", at a distance from Earth of 9.64 AU and $L_{s}$ of 41.86° (local northern spring). The spectral range of the observation was 1151-1160 cm$^{-1}$, the spectral resolving power $R=\lambda/\Delta \lambda \simeq80,000$, and the spectral pixel scale was R=300,000, equivalent to a pixel width in wavenumbers of 0.0038 cm$^{-1}$.  Data were retrieved using a slit aligned to celestial north/south that was initially pointed at the centre of Saturn, then offset north 25" to a sky position where the flatfield and five blank sky observations were taken for calibration and sky subtraction.  The slit was then moved just east of Saturn and stepped in 0.7" steps westward across Saturn’s disk until falling off on the opposite side.  The 0.7" step size is half the 1.4" slit width, producing Nyquist-sampled maps of Saturn’s disk.  We integrated for 2 sec for each step of the scan and coadded 64 separate scans giving a final integration time of 128 secs per pixel in the summed datacube.  Each pixel along the 7" slit length covers 0.36".  The final scan map provides a spatially resolved spectral data cube of Saturn. Other than the limb of the planet, no spatial structure is apparent in the image, which is formed by summing across all the measured wavenumbers (Figure 1). The Doppler shift and emission angle of every pixel of the data cube were calculated using custom software written for TEXES planetary observations by Greathouse, which utilizes the NAIF toolkit (Acton et al., 1996).  Only pixels with an emission angle less than 45° (in order to eliminate regions of limb brightening/darkening) were used in the following analysis so as to remove the influence of limb brightening and darkening.
\begin{figure*}
\begin{centering}
\centerline{\includegraphics[angle=0,scale=.4]{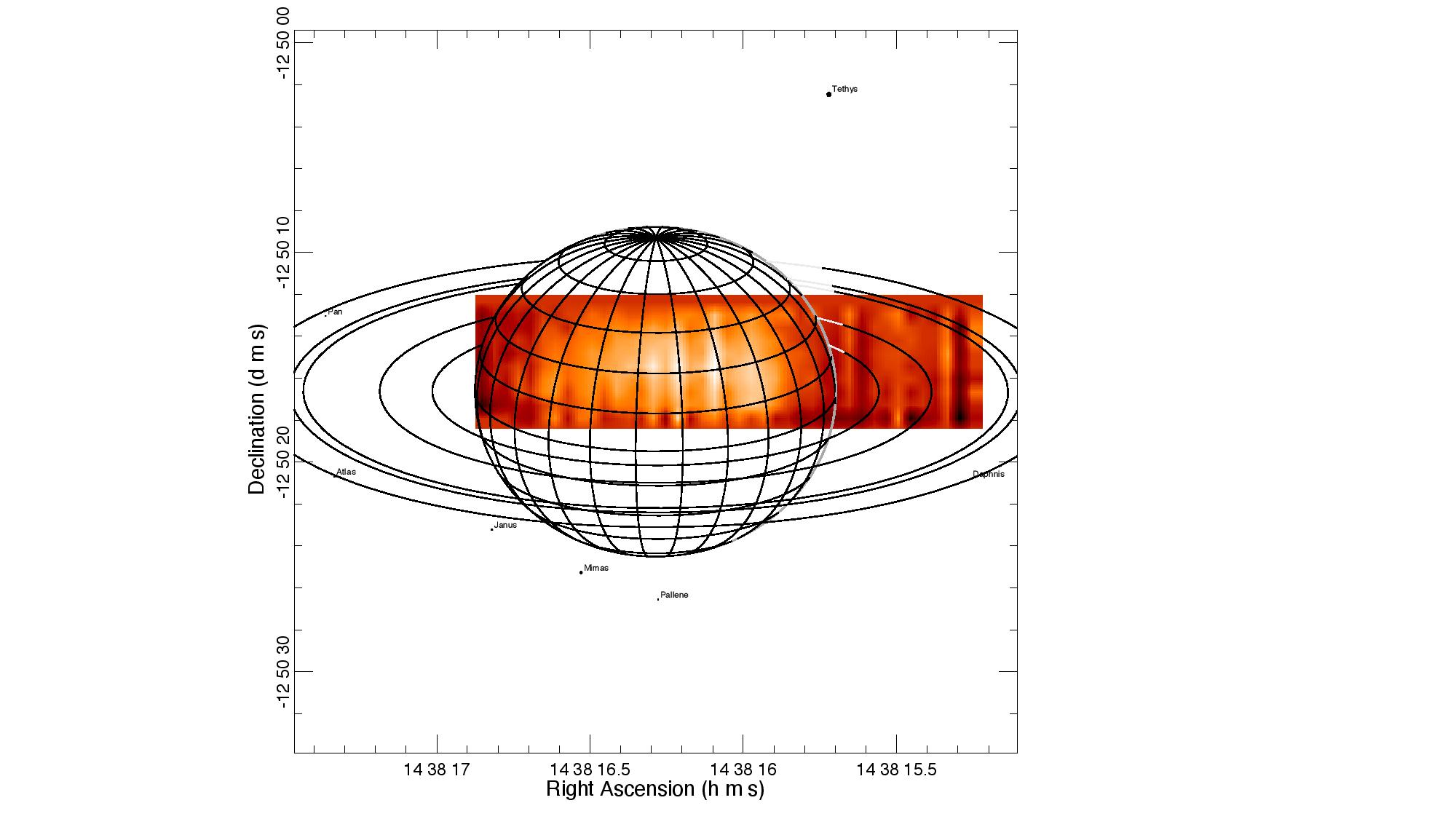}}
\caption{Emission of each mapped pixel shown here with an overlaying wire frame model of Saturn from the midpoint time of the observations. The latitude range along the central meridian is from approximately 10° to 50° north. The background regions (shown in black) do overlap with the rings, however the emission of the rings is undetectable in the wavelength range of this observation.}
\label{thermal_images}
\end{centering}
\label{figure:1}
\end{figure*}

\begin{figure*}
\begin{centering}
\centerline{\includegraphics[angle=0,scale=.525]{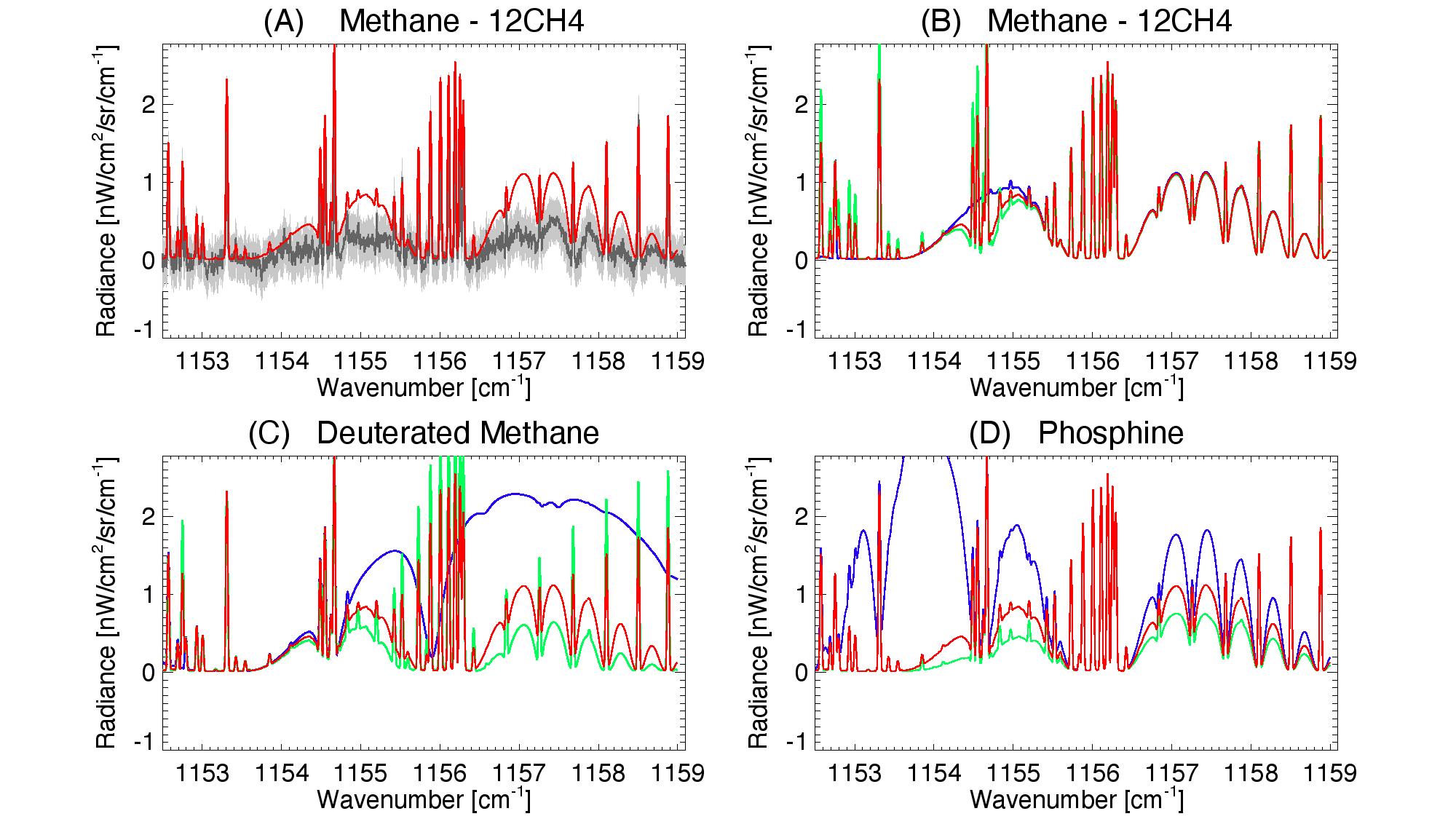}}
\caption{Model spectra showing the influence of varying the abundances of different species. \emph{Panel} a: model spectrum (red solid line) with nominal levels of three methane isotopologues and phosphine, overlaying the observed spectra shifted down by $-0.21 \times 10^{-9} $ W cm$^{-2} $str$^{-1} $cm$^{-1}$  (dark grey solid line) and its associated uncertainties (grey shaded region). \emph{Panel} b: effect of the CH$_4$ population on this spectral range, and similiary for \emph{Panel} c: deuterated methane isotope CH$_3$D and \emph{Panel} d: the tropospheric phosphine (PH$_3$). For each of these three panels (b-d), the red profile is the forward model using a nominal level of each species (as per the model of Moses et al., 2000 and Saturn priors from Fletcher et al., 2009), the green line is a forward model with twice the respective nominal level of each species identified in the title, the blue line is the forward model where the volume mixing ratio (VMR) is set to zero at all altitudes of the identified species.}
\label{thermal_images}
\end{centering}
\label{figure:2}
\end{figure*}

\subsection{Spectral Contributions}

The emission for each wavenumber was averaged across all planetary longitudes and latitudes in this region to improve the signal to noise ratio (S/N). The resulting spectra were then analysed using the non-linear optimal estimator for multivariate spectral analysis code (NEMESIS, Irwin et al., 2008) to fit the spectrum (Figure 2); a more detailed description of NEMESIS can be found in Section 3. Figure 2 compares the averaged TEXES spectrum to three forward-modelled spectra that demonstrate the effects of the methane isotopologues and phosphine on the model spectra. There are many different species that contribute to Saturn’s mid-infrared spectra, either through emission or absorption. From testing the isotopologue 13CH$_4$ it was evident that its contribution to this spectral range was negligible compared to the inherent noise of the data, we therefore can ignore its variation when retrieving the D/H ratio. Though PH$_3$ does have a significant influence on this spectral range (Figure 2), retrieval testing showed that its influence could not be disentangled from that of temperature. Indeed, when PH$_3$ was varied as a free parameter in our spectral inversions, the best fits had the PH$_3$ volume mixing ratio (VMR) largely unchanged. Hence, we assume the PH$_3$ VMR can remain constant for all subsequent retrievals.

\subsection{Radiometric Calibration Uncertainties}

At the outset of the analysis, we discovered that the TEXES spectra contained a number of regions where the calibrated brightness was systematically brighter where our emission model suggests the emission should be close to zero, thus signifying an issue with the radiometric calibration. Indeed, this prevented NEMESIS from fitting a physically plausible temperature structure. To counter this, a standard radiance offset was added to the measured spectra. This value was derived by using the mean difference between the measurement and the forward model to match the continuum in the spectral regions 1153-1154 cm$^{-1}$ and 1159-1160 cm$^{-1}$. A range of values around this number were tested to find the retrieval that gave the lowest $\chi^{2}$ fit of the model to the observed spectra as well as constraining the sensitivity of our D/H result to this radiometric uncertainty. Given that both the methane and deuterated methane lines are measured within the same spectral setting, this systematic offset does not significantly influence the result. The final radiance offset used was  $-0.21 \times 10^{-9} $ W cm$^{-2} $str$^{-1} $cm$^{-1}$ as this yielded both the smallest $\chi^{2}$  value and a realistic temperature retrieval.



\section{Analysis}
\label{Analysis}

NEMESIS (Irwin et al. 2008) is a radiative-transfer and spectral-retrieval code that has been used extensively to measure temperature and composition in giant planet atmospheres. This tool performs a spectral inversion using an optimal estimation approach (Rodgers et al. 2000), thereby minimising a two-term cost function that is comprised of the residual fit to the spectral data and our prior knowledge of the atmospheric state; a priori estimates of atmospheric temperature and the volume mixing ratios of different species as a function of altitude. These a priori estimates are used to constrain the retrieval process by ensuring that the final solution retains physically realistic atmospheric profiles. Sources of spectral line data are identical to those tabulated in Fletcher et al. (2018) and were used to generate k-distributions on a 0.0144 cm$^{-1}$ grid. Spectral line data for methane and deuterated methane came from Brown et al. (2003) and their temperature dependence from Margolis et al. (1993).

A reference atmosphere was used to create the a priori estimates for the atmospheric temperature as well as mole fractions of methane and its isotopologues. The a priori temperature estimate was taken from a model based on the time-series of Cassini Composite Infrared Spectrometer (CIRS) measurements produced by Fletcher et al. (2017). The time series was interpolated to the February 2013 date of the TEXES observations to find a prior for the T(p). The temperatures reported by Fletcher et al. (2017) did not cover the entire pressure range that we require for our NEMESIS analysis (1 microbar to 10 bar), hence we blended this with the temperature model of Moses et al. (2000). A smoothing function was applied to this temperature profile to reduce any discontinuities between the CIRS model and the Moses model temperatures.

The vertical shape of the a priori volume mixing ratio profile (VMR) of CH$_4$ was acquired from the Moses et al. (2000) model of Saturn’s atmosphere with the base level scaled to the CIRS-derived mole fraction of $4.7 \pm 0.2 \times 10^{-3}$ from Fletcher et al. (2009). CH$_{4}$ is predicted to be well mixed throughout the atmosphere of Saturn (Flasar et al., 2005) and therefore serves as an accurate sensor for measuring atmospheric temperature. CH$_{3}$D is expected to be a consistent fraction of the CH$_{4}$ population as there are no production or loss mechanisms within Saturn’s atmosphere that are currently known. For the CH$_{3}$D vertical VMR profile, the a priori estimate was set to be a fraction of the CH$_{4}$ profile in the same manner as Fletcher et al. (2009b) using equation 1.
\begin{equation}
    CH_{3}D(VMR)=4 \times CH_{4}(VMR) \times 1.7\times10^{-5}.
\label{equation1}
\end{equation}

The factor of $1.7 \times 10^{-5}$ was the D/H ratio in hydrogen used in the a priori set up of Fletcher et al. (2009b), it has been multiplied by 4 here to convert to the CH$_{3}$D abundance (see equation 2 in Section 4) assuming a fractionation factor of 1. During retrievals, the VMR profile of CH$_{3}$D was allowed to scale by a constant factor to preserve the shape of the distribution, given that the CH$_{3}$D population must be scaled at each level as a uniform proportion of the CH$_{4}$ population.

Using these a priori estimates, we simultaneously retrieved the temperature and CH$_{3}$D VMR over a selection of CH$_{4}$ emission lines (1152.5-1152.73 cm$^{-1}$, 1153.2-1153.6cm$^{-1}$ and 1154.0-1155.0 cm$^{-1}$) and CH$_{3}$D emission lines (between 1155.8 and 1156.15 cm$^{-1}$). Figure 3 shows the resultant spectral fit (in red) of the CH$_{4}$ and CH$_{3}$D emission lines as well as their surrounding continuum. The uncertainty in the spectral radiance was drawn from the standard deviation of background noise sampled from the off-planet regions of the TEXES spectral cube. This uncertainty was increased for regions of low telluric transmission so that the retrievals would be weighted towards clearer sky spectral windows, thereby reducing the effects of terrestrial contamination. Figure 4 shows the retrieved temperature and CH$_{3}$D VMR altitudinal profiles. Both the troposphere and stratosphere are cooler than the CIRS model would suggest for this time period. The VMR of CH$_{3}$D is a factor of approximately 0.94 of the a priori estimate (Figure 4 (b)).

\begin{figure*}
\begin{centering}
\centerline{\includegraphics[angle=0,scale=.625]{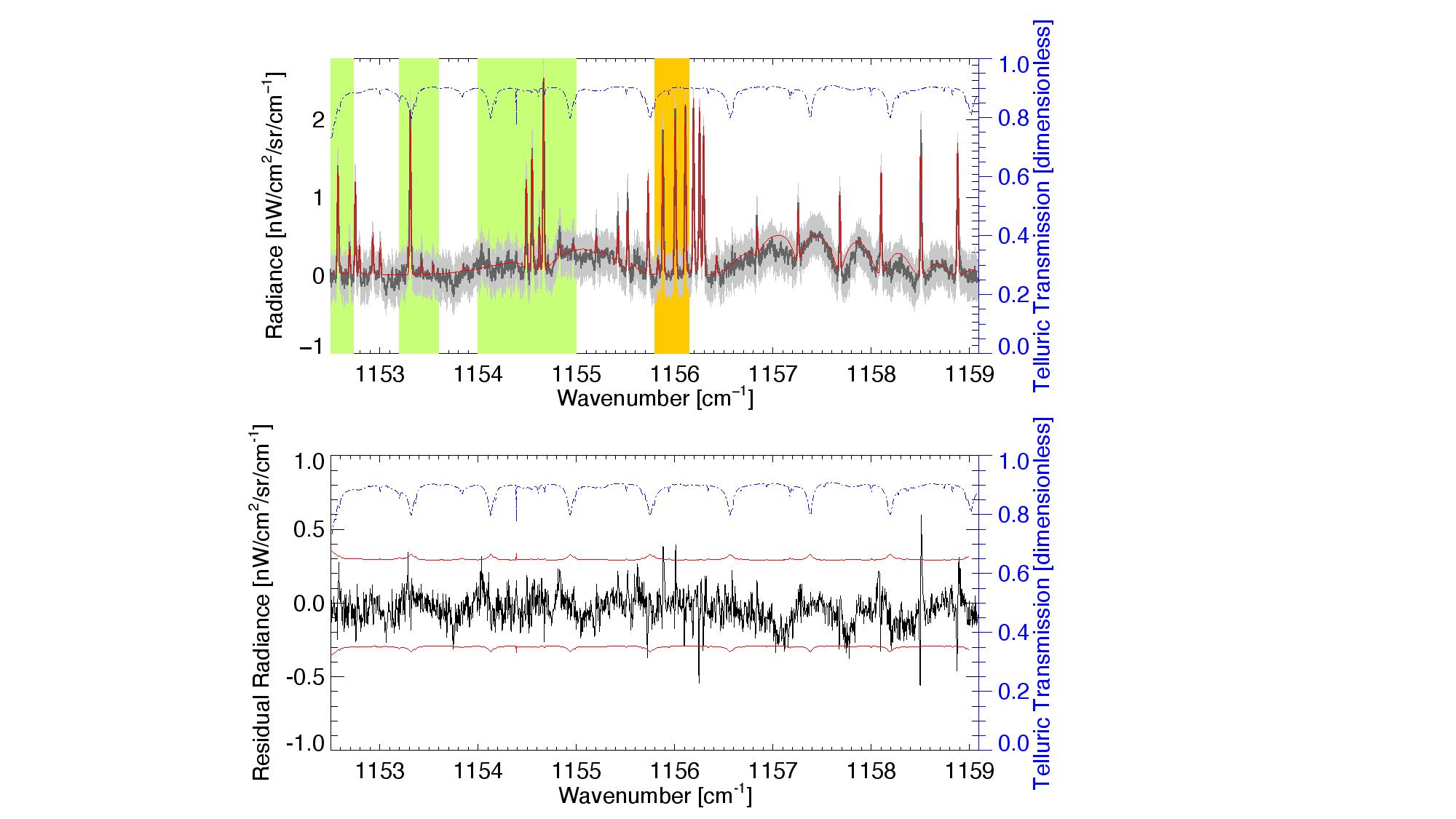}}

\caption{(top) spectral retrieval of the CH$_4$ emission lines and the CH$_3$D emission lines (red line), the spectra from the observation (black solid line), the shaded grey region is the uncertainty range associated with the observed spectra, the telluric absorption (blue dot-dashed line) in this spectral range. The spectral ranges used in the retrieval are highlighted for CH$_4$ (green) and CH$_3$D (orange). (bottom) residual difference between the spectral retrieval and the spectra from the observation (black), the observation uncertainty (red solid line) is shown for each wavelength as well as the telluric absorption (blue dot-dashed line).}

\end{centering}
\label{figure:3}
\end{figure*}

\begin{figure*}
\begin{centering}

\centerline{\includegraphics[angle=0,scale=.45]{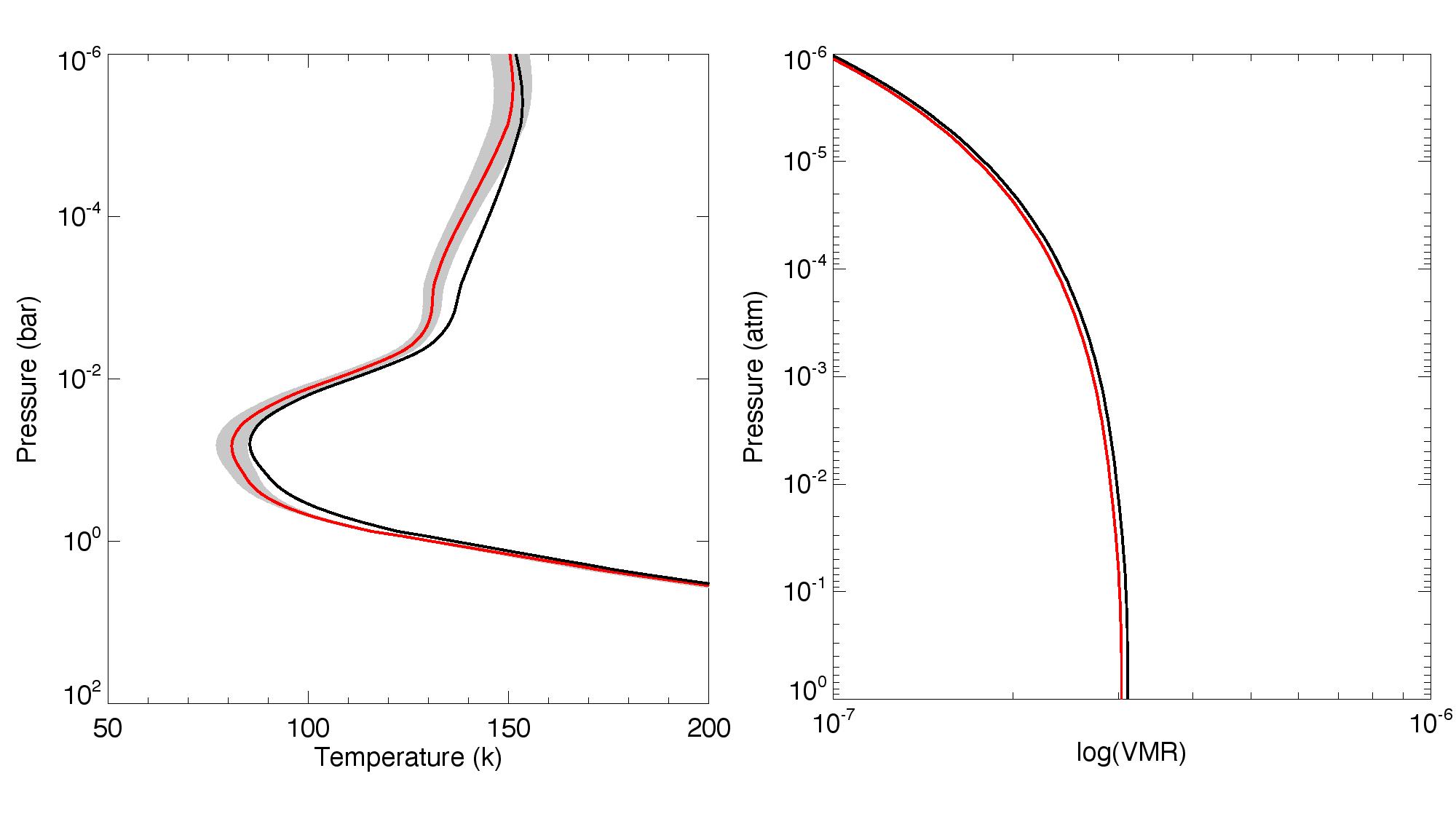}}

\caption{\emph{Left}: retrieved vertical temperature profile T(p) is shown in red and the shaded grey region is the associated uncertainty, the a priori T(p) estimate is shown in black. \emph{Right}: retrieved vertical VMR profile of CH$_3$D is shown in red alongside it’s a priori estimate, which is shown in black.}

\end{centering}
\label{figure:4}
\end{figure*}

We retrieve a CH$_{3}$D abundance at the 1-millibar level of $0.30_{-0.05}^{+0.04}$ ppm, which is close to the a priori estimate. Taking the ratio of the retrieved CH$_{3}$D vertical VMR and the CH$_{4}$ VMR profile, we estimate Saturn’s (D/H)$_{CH_{4}}$ $= 1.65 \times 10^{-5}$, with a full uncertainty analysis in the following Sections. We note that this value is considered to be representative of Saturn’s stratosphere, because this is where we have an independent constraint on the atmospheric temperatures from the nearby CH$_4$ emission lines. While we do have sensitivity to tropospheric CH$_{3}$D, we do not have a way of independently constraining tropospheric temperatures (not to mention the influence of tropospheric phosphine).  Hence, we assume that the value derived in the stratosphere is relevant over the entire atmospheric column.
In the following sections we conduct analysis for the degree of uncertainty in this D/H ratio. We test the sensitivity of our fit to various input parameters; the uncertainty in the temperature fit (Section 3.1) and the uncertainty caused by the radiance offset (Section 3.2).

\subsection{Uncertainty in the temperature fit}

To assess the sensitivity of the spectral fits to the derived D/H ratio, we fixed the CH$_{3}$D abundance to values that are between 50-150\% of the value derived above, and reran the temperature retrieval, comparing the goodness-of-fit to the TEXES spectrum in each case. The $\Delta \chi^{2}$ (the difference of the non-reduced $\chi^{2}$ from the minimum value) for each fit is shown in Figure 6, along with a selected range of retrieved temperature profiles, indicating that a different D/H ratio can yield a slightly different (but still physically plausible) temperature structure. The first confidence interval $\sigma _{1}$ can be measured by interpolating between these points with a quadratic polynomial and examining the point at which these spectral fits deviate from the measurements by a $\Delta \chi^{2}$  of 1. Similarly, the confidence intervals $\sigma _{2}$ and $\sigma _{3}$ were estimated from deviations with a $\Delta \chi^{2}$  of 4 and 9 respectively. It is apparent from Figure 5 that the change to the D/H ratio does not affect the $\chi^{2}$  value in a symmetrical fashion about the minimum. As such, two different quadratic fits were used for the D/H values less than (shown in blue) and greater than (shown in red) the initial retrieval. The confidence intervals for this fit are $\sigma _{1}= -0.20/+0.27 \times 10^{-5}$  , $\sigma _{2}= -0.38/+0.51 \times 10^{-5}$ and $\sigma _{3}= -0.57/+0.93 \times 10^{-5}$. The first confidence interval corresponds to an uncertainty on the D/H ratio of approximately 12\%, suggesting that the temperature retrieval is a significant source of uncertainty in this analysis of D/H. Using the first confidence interval, the D/H in methane with uncertainties is therefore given as $1.65_{-0.20}^{+0.27} \times 10^{-5}$. However, this does not capture the additional systematic uncertainties related to the TEXES calibration, as explained below.

\begin{figure*}
\begin{centering}

\centerline{\includegraphics[angle=0,scale=.45]{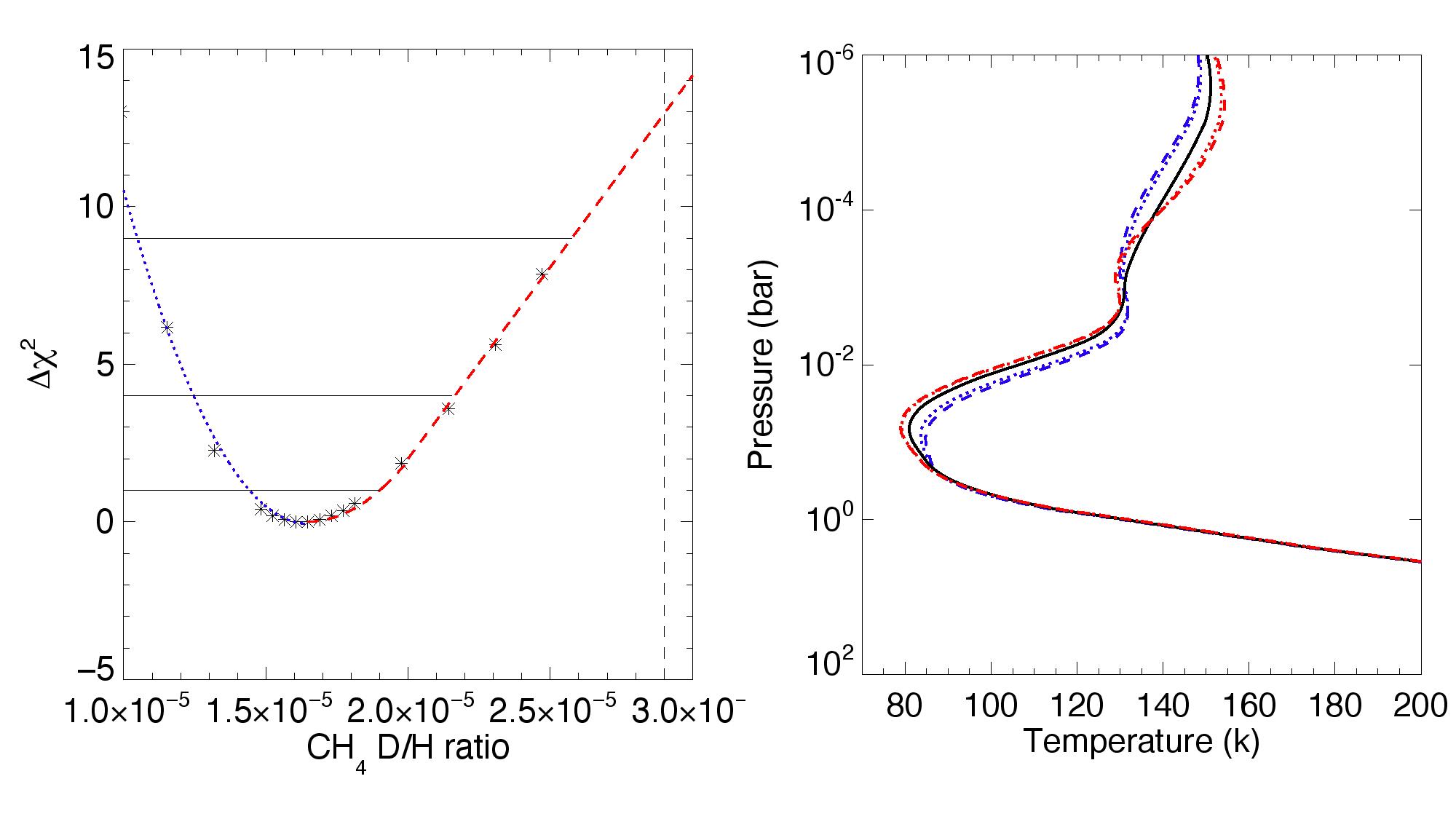}}

\caption{\emph{Left}: $\chi^{2}$  values (compared to the best fit) of each temperature retrieval for the range of fixed D/H ratios. Measured values are marked by the black asterisks, the blue dotted line marks the quadratic fit to the D/H ratios less than $1.65 \times 10^{-5}$ and the dashed red line marks the quadratic fit of the D/H ratios that are greater than $1.65 \times 10^{-5}$. The minimum $\chi^{2}$ of 84.487 has been subtracted from all values so that this plot can show $\Delta \chi^{2}$. The horizontal black lines delineate the $\Delta \chi^{2}$ of 1, 4 and 9 which correspond to $\sigma_1$, $\sigma_2$  and $\sigma_3$ respectively. The vertical dotted line marks the inferred value of methane D/H $2.9 \times 10^{-5}$ converted from the hydrogen proto-solar D/H of Geiss and Gloeckler (1998) using a fractionation factor of 1.34, demonstrating that the proto-solar quantity of deuterated methane provides a significantly worse fit using our data and retrieval. \emph{Right}: the temperature profiles associated with CH$_3$D abundances set to 50\% (blue dashed line), 60\% (blue dotted line), nominal levels (solid black line), 140\% (red dotted line) and 150\% (red dashed line). We note that the nominal level is in the centre of these adjusted temperature retrievals, this indicates that the black profile is the best temperature fit for these data.}

\end{centering}
\label{figure:5}
\end{figure*}

\subsection{Uncertainty esimation of radiance offset}
To test the sensitivity of the retrieval to the systematic radiance offset described in Section 2.2 a series of NEMESIS retrievals were performed for a range of radiance offsets applied to the spectra. A range of radiance offsets from $-1.3$ to $-3.0 \times 10^{-9} $ W cm$^{-2} $str$^{-1} $cm$^{-1}$ were each tested with a retrieval of both the temperature and CH$_{3}$D VMR. Using a least-squares fitting technique, the $\chi^{2}$  values of each of these retrievals were assessed, as well as the D/H ratio. The minimum $\chi^{2}$ was subtracted from all measured $\chi^{2}$  values, such that the resultant contour plot shows $\Delta \chi^{2}$ (Figure 6). The first three confidence intervals are marked respectively, $\sigma _{1}$, $\sigma _{2}$  and $\sigma _{3}$, which correspond to a $\Delta \chi^{2}$  of 1, 4 and 9. The confidence intervals for this fit are $\sigma _{1}= -0.39/+0.18  \times 10^{-9}$, $\sigma_{2}= -0.89/+0.35 \times 10^{-9}$ and $\sigma _{3}= -1.37/+0.55×10^{-9}  $ W cm$^{-2} $str$^{-1} $cm$^{-1}$. The first confidence interval amounts to a change in the D/H of $ -0.047/+0.013 \times 10^{-5}$ (Fig. 6) which corresponds to an uncertainty of approximately 1-3\% in the D/H ratio. This highlights the benefit of measuring both the methane and deuterated methane lines together in the same TEXES spectral setting, rendering them largely immune to systematic offsets. This indicates that the radiance offset has a smaller influence on the measurement of D/H and the quality of the spectral fits than the temperature-fitting process of Section 3.1.

\begin{figure*}
\begin{centering}

\centerline{\includegraphics[angle=0,scale=.45]{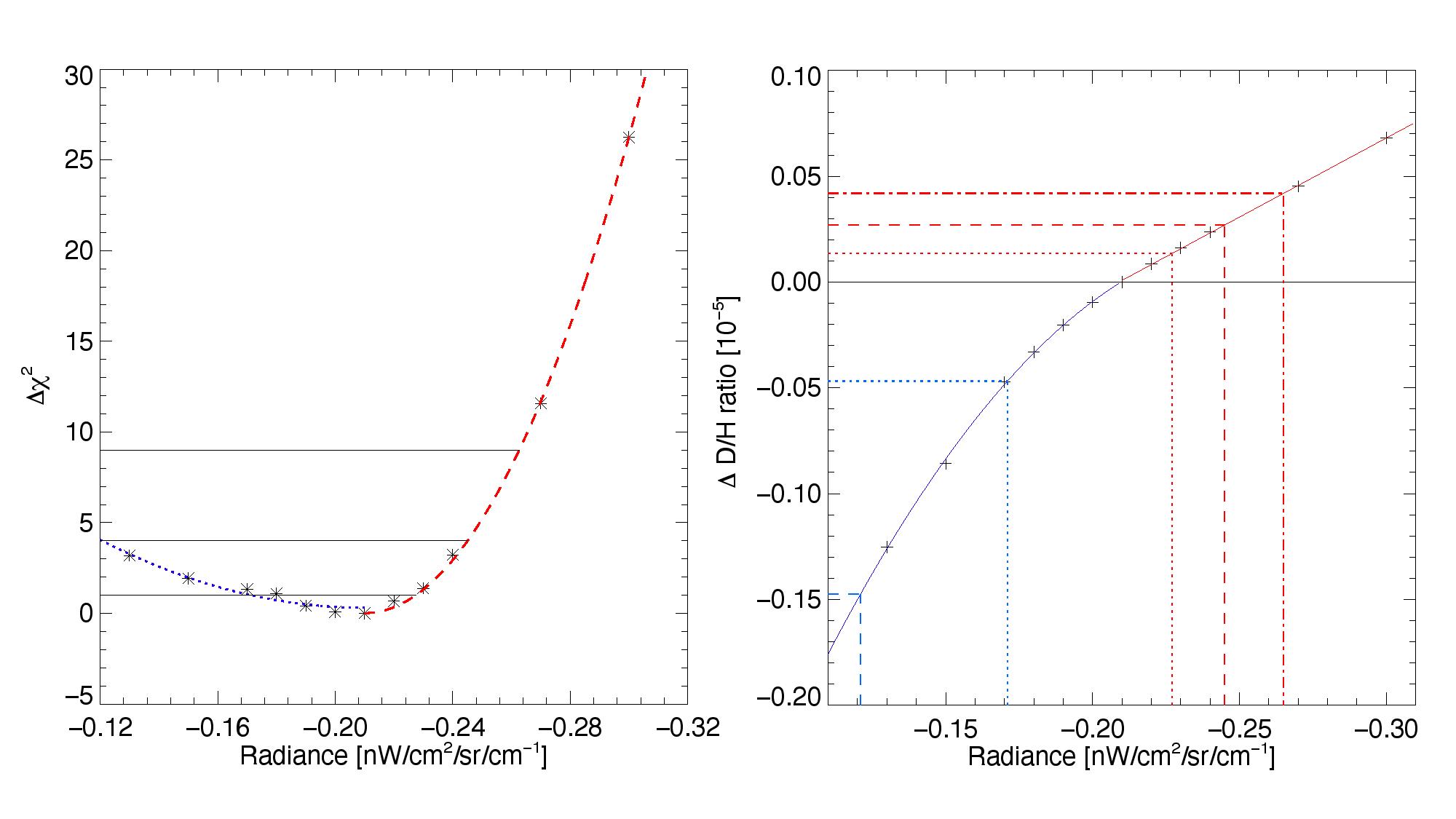}}

\caption{\emph{Left} Non-reduced $\Delta \chi^{2}$ values obtained from NEMESIS retrievals of (D/H)$_{CH_{4}}$ and temperature as a function of radiance offset. The three horizontal lines represent a $\Delta \chi^{2}$ of 1, 4 and 9, which are equivalent to the $\sigma_1$, $\sigma_2$ and $\sigma_3$ levels, respectively. The dotted blue and dashed red lines mark the quadratic fits to the relevant regions. \emph{Right} Change in the D/H ratio as a function of radiance offset. Derived values are marked by black crosses for each radiance offset. Quadratic fits to the radiance are marked for the same radiance offsets as in the left panel. For each of the confidence intervals, the radiance shift and its corresponding change in D/H is marked by $\sigma_1$ dotted (red and blue), $\sigma_2$ dashed (red and blue) and $\sigma_3$ dot-dashed (red only).}

\end{centering}
\label{figure:6}
\end{figure*}

\section{Discussion}
\label{discuss}
We measure Saturn’s D/H ratio for methane to be $1.65_{-0.21}^{+0.27} \times 10^{-5}$, with the stated 1-sigma uncertainty representing the contribution of the NEMESIS fit, the radiance offset and the uncertainty in the CH$_{4}$ abundance (adding these uncertainties in quadrature). The uncertainty due to the temperature retrieval is by far the most significant influence on the D/H ratio at ~12\%. In contrast, systematic errors in the TEXES radiance calibration appear to only have a minimal influence (1-3\%). These uncertainties could be further constrained with new observations of Saturn at the same resolution but with a broader wavelength coverage, thereby improving the S/N ratio and sensitivity to stratospheric methane lines, which in turn would provide a more constrained temperature retrieval. 

Our D/H measurement is in agreement with the methane D/H of $1.6 (\pm 0.2) \times 10^{-5}$ derived from the Cassini/CIRS observations of CH$_{3}$D by Fletcher et al. (2009b), but not with the estimate of $2.85 (\pm 0.2) \times 10^{-5}$ that was estimated by converting the D/H derived from measurements of HD by Pierel et al (2017).  Our D/H in methane is within the uncertainties of the result of $1.7 (\pm1.1) \times 10^{-5}$ determined from high-resolution 5 $\mu$m spectroscopy by Noll and Larson (1991), and is at the lower end of the ISO/SWS determination of $2.0_{-0.7}^{+1.4} \times 10^{-5}$  by Lellouch et al. (2001).  It appears to be too low to be consistent with the estimate of 2.4 (±0.5) $\times 10^{-5}$ in methane from Bézard et al. (2003), who used the same TEXES instrument and spectral region in 2000, but we have been unable to reproduce their result.

\subsection{Deuterium fractionation}

The D/H ratio of methane and hydrogen are related to each other as shown in equation 2. Prior measurements of Saturn’s D/H ratio have utilised the isotopic enrichment factor f which relates the distribution of the deuterium isotope in hydrogen-containing molecular species.
\begin{equation}
    f = \frac{(D/H)_{CH_{4}}}{(D/H)_{CH_{3}D}}.
    \label{equation2}
\end{equation}

Lellouch et al. (2001) used a fractionation factor of $1.34 \pm 0.19$ taken from the average of prior assessments using both laboratory analysis (Lécluse et al., 1996) and model analysis (Smith et al., 1996, 1998). This fractionation factor was most recently used in the work of Pierel et al. (2017) to convert their (D/H)$_{H_{2}}$ to a D/H for methane of $2.85 \times 10^{-5}$. Working in the opposite direction, if we apply this fractionation factor to our (D/H)$_{CH_{4}}$ we estimate a hydrogen D/H ratio of $1.23_{-0.23}^{+0.27} \times 10^{-5}$ which is significantly below the uncertainty ranges of the estimated proto-solar values $2.1 (\pm 0.5) \times 10^{-5}$ from Geiss and Gloeckler (1998).

Surprisingly, our (D/H)$_{H_{2}}$ estimated from our methane analysis is not within the uncertainty range of the CIRS analysis from Pierel et al. (2017) which yielded a hydrogen-derived D/H value of $2.10 (\pm0.13) \times 10^{-5}$. Indeed, if we compare our D/H in methane with the CIRS-derived D/H in hydrogen, we would need an entirely different fractionation factor of approximately 0.79.  Conversely, our value is within the uncertainty range of the ISO-SWS measurement (D/H)$_{H_{2}}$ of  $1.85_{-0.60}^{+0.85} \times 10^{-5}$  for Saturn from Lellouch et al (2001).  NEMESIS re-analysis of the same ISO-SWS results gave a  (D/H)$_{H_{2}}$ value of $1.62 (\pm0.61) \times 10^{-5}$ (see Table 5 of Pierel et al., 2017), which is even closer to our estimated (D/H)$_{H_{2}}$ from the TEXES measurements. Note that taking this latter (D/H)$_{H_{2}}$ and comparing it to our refined value of (D/H)$_{CH_{4}}$ would imply a fractionation factor close to unity.  Laboratory work to reassess the fractionation factor under Saturn’s stratospheric conditions would be welcome.

We initially suspected that the significant scatter in the D/H results might be a consequence of the differing spectral resolutions used in different studies.  However, Griffin et al. (1996) found a higher hydrogen D/H of $2.3_{-0.8}^{+1.2}\times10^{-5}$  when using the ISO – LWS which has a maximum resolving power of R=9,000. From this we infer that the variation in D/H measurements is unlikely to be caused by variation in spectral resolution, as the high-resolution spectra of TEXES appears to confirm lower-resolution measurements from CIRS (Fletcher et al., 2009b). An alternative source of this variation could be the spectroscopic parameters used. In this work we use the HITRAN line database, however Lellouch et al. (2001) used the GEISA 1997 data bank. Further testing is required to see if this would make an appreciable difference. In summary, our new methane-derived stratospheric D/H value is consistent with one derived from Cassini CIRS data, but both results are generally lower than what would be expected from the D/H measured independently in hydrogen at tropospheric depths. This raises the intriguing possibility that the deuterium fractionation might vary between the troposphere and stratosphere reservoirs.  However, this inconsistency between different spectral regions might also be a consequence of discrepancies in the spectroscopic line data for HD, and a re-analysis of CIRS and ISO HD observations might be required with a range of different scale factors applied to the rotational line intensities, to see if there is some way to make the two spectral ranges consistent.  It may also be that the CH$_{3}$D measurements suffer from fewer systematic effects given the spectral close nature of the CH$_{3}$D and CH$_{4}$ emissions compared to the much more spectrally separated H$_{2}$ and HD lines.

\subsection{Comparisons to Jupiter and Neptune}

Direct comparisons of our D/H in Saturnian stratospheric methane to published D/H in Jovian stratospheric methane, $2.2 (\pm0.7) \times 10^{-5}$ (Lellouch et al., 2001) and $2.4 (\pm0.7) \times 10^{-5}$ (Bjoraker et al., 1986), show that our result for Saturn is significantly lower, however still marginally within uncertainty values.  We also note that the Saturnian (D/H)$_{CH_{4}}$ is a factor of 3-6 times smaller than that found in Neptune’s methane ($7.5 (\pm 2.5) \times 10^{-5}$ as deduced in Section 1, based on Fletcher et al., 2010; Irwin et al., 2014).

Our inferred D/H in hydrogen of $1.23_{-0.23}^{+0.27} \times 10^{-5}$  is substantially lower and outside of the uncertainty margins of Jupiter’s hydrogen D/H from ISO of $2.4 (\pm 0.4) \times 10^{-5}$ (Lellouch et al., 2001), and the in situ Galileo-probe measurement $2.6 (\pm 0.7) \times 10^{-5}$ (Mahaffy et al., 1998). Given the margin of uncertainty on the Jupiter and Saturn measurements, as well as the proposed values for the proto-solar value of D/H, this comparison indicates that the methane D/H ratio in Saturn’s stratosphere is definitively lower than that of Jupiter and the proto-solar value. It remains a possibility that some sort of fractionation process, depleting deuterium in the observable atmosphere of Saturn but not Jupiter, may be playing a role. For example, Dobrijevic and Loison (2018) have previously found photochemical fractionation of N isotope-bearing molecules in Titan’s atmosphere.  Contamination of Saturn’s stratosphere via exogenic materials with lower D/H ratios could also be a possibility, although this is unlikely given the high D/H observed in cometary ices. Whilst we cannot definitively say the same processes are responsible for our measured D/H of Saturn, perhaps the D/H measured in stratospheric methane emission is not wholly representative of Saturn’s bulk composition due to some unidentified chemical process similar to Dobrijevic and Loison (2018). The tropospheric D/H measured in hydrogen by, for example, Pierel et al. (2017) might therefore be the better estimate pertinent to the study of planetary origins.  

A possible method to reconcile this would be to measure the D/H of Jupiter and Saturn using the same observational and analytical techniques, thereby allowing a direct comparison to be made – the James Webb Space Telescope (Norwood et al., 2016) might offer such an opportunity in the CH$_{3}$D and CH$_4$ emission regions, but the rotational HD features will be out of reach.

\section{Conclusions}
\label{conclusion}
Infrared spectra from the TEXES instrument have been analysed over the range of 8.62 to 8.68 $\mu$m ($1151-1160 $cm$^{-1}$) in order to determine the D/H ratio in Saturn’s stratospheric methane. This spectral range features emission lines from both methane and its deuterated isotopologue and was used to retrieve the atmospheric temperature and deuterated methane volume mixing ratio simultaneously as a function of altitude. We find a methane D/H of $1.65_{-0.21}^{+0.27} \pm 10^{-5}$, which is in good agreement with previous D/H measurements in methane in the same spectral range from Cassini CIRS (Fletcher et al., 2009b). However, this stratospheric value is considerably smaller than that found for Jupiter and, if we employ a factor to account for fractionation in different molecules, it would predict a D/H in hydrogen that is significantly smaller than previous ISO and Cassini measurements of far-infrared HD lines.

We note that estimations of the hydrogen D/H on Saturn are consistently lower than the protosolar value of D/H (estimated at $2.1 (\pm 0.5) \times 10^{-5}$ – Geiss and Gloeckler, 1998) and are actually closer to those values found today in the Local Interstellar Medium - D/H in hydrogen of $1.51 (\pm0.1) \times 10^{-5}$ (Linsky, 1998, Sahu et al., 1999). No explanation for this disparity has been found so far and we recognise that it is difficult to imagine a mechanism which would deplete deuterium at Saturn. It is possible that the stratospheric value is not representative of Saturn’s bulk, either via some unidentified chemical fractionation process, or maybe by contamination from unidentified exogenic material with a D/H value lower than proto-solar levels. Spatially resolved mapping of variations in the D/H ratio might help elucidate these possibilities, but further progress may require in situ sampling of the D/H ratio via atmospheric entry probes for Saturn such as the HERA Saturn entry probe proposed in Mousis et al. (2016).

\section*{Acknowledgments}
\label{Acknowledgments}
The Leicester authors were supported by a European Research Council Consolidator Grant, under the European Union’s Horizon 2020 research and innovation framework, grant number 723890.  LNF was also supported by a Royal Society Research Fellowship at the University of Leicester.  The TEXES observations were acquired under programme ID 2013A-045 at NASA’s Infrared Telescope Facility, where Greathouse and Fletcher were visiting astronomers. Greathouse and Orton were supported by NASA with funds distributed, respectively, to the Southwest Research Institute and to the Jet Propulsion Laboratory, California Institute of Technology. The Infrared Telescope Facility is operated by the University of Hawaii under contract NNH14CK55B with the National Aeronautics and Space Administration.  We recognise the cultural and religious significance of Maunakea to the Hawaiian people and are privileged to have been able to observe from this mountain.

\section*{References}
\label{References}

    Acton, C. H. 1996, Planetary and Space Science, 44(1), 65-70.

Bézard, B., Greathouse, T., Lacy, J., Richter, M., \& Griffith, C. A. 2003, BAAS, 35, 1017

Bjoraker, G. L., Larson, H. P., \& Kunde, V. G. 1986, Icar, 66, 579

Brown, L. R., Benner, D.C., Champion, J.P, et al. 2003, Journal of Quantitative Spectroscopy and Radiative Transfer, 82, 219–238

Dobrijevic, M., \& Loison, J.C., 2018, Icarus 307, 371

Fegley, B. \& Prinn, R.G., 1988, Atrophys J., 326, 490-508

Ferlet, R., \& Lemoine, M. 1998, SSRv, 84, 297

Fletcher, L. N., Orton, G. S., Teanby, N. A., Irwin, P. G. J., \& Bjoraker, G. L. 2009b, Icarus, 199, 351

Fletcher, L.N, Drossart, P., Burgdorf, M. et al. (2010). A\&A, 514, A17

Fletcher, L. N., Greathouse, T. K., Orton, G. S., et al. 2014, Icarus, 238, 170

Fletcher, L.N., Guerlet, S., Orton, G., (2017), Nature Astronomy, 1, 765-770

Fletcher, L.N., Greathouse, T.K., Guerlet, S., (2018), Saturn in the 21st Century, (Cambridge: Cambridge University Press), doi:10.1017/9781316227220

Feuchtgruber, H., Lellouch, E., Bezard, B., Encrenaz, T, et al., 1999, A\&A, 341, L17-L21

Feuchtgruber, H., Lellouch, E., Orton, G., et al. 2013, A\&A, 551, A126

Flasar, F. M., Achterberg, R. K., Conrath, B. J., et al. 2005, Science, 307, 1247:1251.

Fouchet, T., Greathouse, T.K., Spiga, A., et al. 2016, Icarus, 277, 196-214

Geiss, J., \& Gloecker, G. 1998, SSRv, 84, 239

Greathouse, T.K., Lacy, J.H., Bézard, B., et al. 2006, Icarus, 181, 266–271

Griffin, M. J., Naylor, D. A., Davis, G. R., et al. 1996, A\&A, 315, 389

Hersant, F., Gautier, D., \& Huré, J. 2001, ApJ, 554, 391

Irwin, P. G. J., Teanby, N. A., de Kok, R., et al. 2008, JQSRT, 109, 1136

Irwin P.G.J., Lellouch, E., de Bergh, C., et al. 2014, Icarus, 227, 37-48

Lacy, J. H., Richter, M. J., Greathouse, T. K., Jaffe, D. T. \& Zhu, Q. 2002, JSTOR, 114(792), 153-168.

Lécluse, C., Robert, F., Gautier, D., \& Guiraud, M. 1996, Planet Space Sci., 44, 1579

Lellouch, E., Bezard, B., Fouchet, T., et al. 2001, A\&A, 370, 610

Linsky, J. L. 1998, Space Sci. Rev., 84, 285

Mahaffy, P. R., Donahue, T. M., Atreya, S. K., et al. 1998, SSRv, 84, 251

Margolis, J. S. 1993, Journal of Quantitative Spectroscopy and Radiative Transfer 50, 431–441 

Moses, J. I., Bezard, B., Lellouch, E., et al. 2000, Icarus, 143, 244

Mousis, O., Atkinson, D.H., Spilker, T. et al. 2016, P\&SS, 130, 80-103

Noll, K. S., \& Larson, H. P. 1991, Icarus, 89, 168

Norwood, J., Moses, J., Fletcher, L.N., et al. 2016, PASP 128, 018005

Owen, T. C., Lutz, B. L., \& de, B. C. 1986, Nature, 320, 244

Pierel, J.D.R., Nixon, C.A., Lellouch, E., et al. 2017, ApJ, 154:178

Press, W., Teukolsky, S., Vetterling, W., Flannery, B. 2007, Numerical Recipes, Cambridge University Press, Third Edition

Rodgers, C.D. 2000, Inverse Methods for Atmospheric Remote Sounding: Theory and Practice. World Scientific

Sahu, M. S., Landsman, W., Bruhweiler, F. C., et al. 1999, ApJ, 523, L159

Smith, M. D., Conrath, B. J., \& Gautier, D. 1996, Icarus, 124, 598

Smith, M. D. 1998, Icarus, 132, 176

\end{document}